\documentclass[prd,showpacs,preprintnumbers,amsmath,amssymb]{revtex4}

\usepackage{graphicx}
\usepackage{dcolumn}
\usepackage{bm}

\begin{document}

\title{The Dark Energy Regulated by Emergent Conformal Symmetry}

\author{Yongsung Yoon}\email{cem@hanyang.ac.kr}
\affiliation{Physics Department, Research Institute for Natural Sciences, Hanyang University, Seoul 133-791, Korea}

\date{\today}

\begin{abstract}
We have found a mechanism which regulates the dark energy in our universe. With an emergent conformal symmetry, the
dark energy density is regulated to the order of a conformal anomaly parameter in the conformally coupled gravity. In
the late time cosmological evolution, we have obtained a set of exact cosmological equations which deviate from the
Friedmann equations significantly. Based on the recent observational cosmic expansion data, it is shown that the dark
energy density is about $1/4$ of the matter density at present, which is quite smaller than determined by General
Relativity. The jerk parameter at present is also determined as a definite value $0.47$.
\end{abstract}  \maketitle

\section{Introduction}

The cosmological constant problem has been a big mystery since the development of General Relativity. There have been
many attempts to cancel or remove the cosmological constant without success. Recently, there have been new astronomical
observations of the late time cosmic acceleration of our universe \cite{ACC1,ACC2} adding another mystery to the
cosmological constant problem: the dark energy density in our present universe is not zero, but as small as
$\rho^{obs}_{\Lambda} \sim (10^{-3}{\rm eV})^{4}$, which is quite close to the matter density at present \cite{I1,I2}.

Thus, on the top of the old puzzle for the dark energy: how the vacuum energy in matter Lagrangian could not contribute
to the dark energy, we are now facing another formidable new puzzle: why the dark energy density is so close to the
matter density at present in our universe. As yet, there has been no successful mechanism to relax the vacuum energy
leaving such a small dark energy density which is so close to the matter density in the present universe.

Before the success of Weinberg-Salam model, the weak interactions were characterized by the dimensional Fermi's
coupling constant, $G_{F} \simeq (300 Gev)^{-2}$. But later the Fermi's coupling constant is replaced with the Higgs
scalar field which seems to be discovered recently \cite{Higgs}. Inspired by the success, it seems plausible to replace
the dimensionful Newton's constant with a new gravity scalar field as Brans-Dicke gravity model \cite{BD1,BD2}, or the
induced gravity models \cite{IGO1,IGO2,IGO3,IGO4,IGO5,IG0,Odint,IG1,IG2,IG3,IG4,IG5,IG6,IG7,IG8,IG9}. Furthermore,
there is no room to conceive any mechanism to resolve the dark energy puzzles in General Relativity. Thus, we consider
a gravity model replacing the Newton's constant with a gravity scalar field, adding a cosmic scalar potential.

It is known that, in any gravity models with a gravity scalar field which can give a flat spacetime solution with a
constant gravity scalar field in vacuum, it is not possible to find an exact cancellation mechanism of the cosmological
constant \cite{No-go}.

However, with an evolving gravity scalar field in the late time universe, we have found a regulation mechanism of dark
energy in the conformally coupled gravity with a uniquely determined cosmic scalar potential, which might have an
origin of a conformal anomaly generated by quantum effects in matter Lagrangian. Even though the conformally coupled
gravity model does not give a flat spacetime for a constant gravity scalar field in vacuum, the non-flatness would be
quite negligible in a non-cosmic scale gravity with the uniquely determined cosmic potential.

In the matter dominated late time universe, we have obtained a set of exact cosmological equations which deviate from
the Friedmann equations significantly. However, in the radiation dominated early universe, we regain the Friedmann
equations of General Relativity having a constant gravity scalar field. Thus, we can say that the conformally coupled
gravity with the logarithmic cosmic potential is a minimal extension of General Relativity to implement the dark energy
regulation mechanism, replacing the dimensionful Newton's constant with the gravity scalar field.

\section{The Conformally Coupled Gravity}

Replacing the Newton's constant $G$ with a gravity scalar field $\Phi$ as $8\pi G=1/\Phi^{2}$, we consider the
following conformally coupled gravity action with a cosmic scalar potential $V(\Phi)$ which is to be determined in the
next section,
\begin{eqnarray}
S= -\int d^{4}x\sqrt{-g}\left[\frac{1}{2}\Phi^{2} R +3(\partial \Phi)^{2}\right] + \int d^{4}x\sqrt{-g} \left({\cal
L}_{m} - V(\Phi) \right), \label{action}
\end{eqnarray}
where ${\cal L}_{m}$ is the matter Lagrangian. The first gravity part of the action (\ref{action}) is invariant under
the local conformal transformation, $g_{\mu\nu}(x) \rightarrow e^{2\sigma(x)} g_{\mu\nu}(x), \quad \Phi(x) \rightarrow
e^{-\sigma(x)} \Phi(x)$, which is the reason why we call the action (\ref{action}) as the conformally coupled gravity.
However, the second matter part of the action (\ref{action}) does not have the conformal symmetry, in general.

We have the equations of motion for the gravity scalar field $\Phi$ and the metric $g_{\mu\nu}$ as follows,
\begin{eqnarray}
6\nabla^{2}\Phi = \Phi R + \frac{\partial V(\Phi)}{\partial \Phi} , \label{S-eq}
\end{eqnarray}
\begin{eqnarray}
\Phi^{2} G_{\mu\nu} - 2\Phi\nabla_{\mu}\nabla_{\nu}\Phi +2g_{\mu\nu}\Phi\nabla^{2}\Phi +4\nabla_{\mu}\Phi
\nabla_{\nu}\Phi - g_{\mu\nu}(\nabla\Phi)^{2} - g_{\mu\nu}V(\Phi) = T^{(m)}_{\mu\nu} , \label{G-eq}
\end{eqnarray}
where $T^{(m)}_{\mu\nu}$ is the matter energy-momentum tensor.

Taking the trace of Eq.(\ref{G-eq}) gives an expression of $\Phi^{2}R$. Plugging it into Eq.(\ref{S-eq}), we have a
characteristic constraint equation of the conformally coupled gravity,
\begin{eqnarray}
T^{(m)} +4V(\Phi) -\Phi\frac{\partial V(\Phi)}{\partial \Phi} = 0 , \label{GS-eq}
\end{eqnarray}
where $T^{(m)}$ is the trace of the matter energy-momentum tensor. The origin of the constraint (\ref{GS-eq}) is that
the gravity scalar field $\Phi$ is not dynamical, but auxiliary in the conformally coupled gravity. Even though the
total classical action (\ref{action}) was not imposed to have a conformal symmetry in general, the consistency in
equations of motion (\ref{S-eq},\ref{G-eq}) requires the conformal symmetry equation (\ref{GS-eq}) in the conformally
coupled gravity \cite{YP2}. Therefore, the conformally coupled gravity (1) describes a system of having an emergent
conformal symmetry, but not having the conformal symmetry in the classical action. \footnote[1]{Replacing the
coefficient $3$ in front of $(\partial\Phi)^{2}$ with a constant $\xi > 6$ in the action (\ref{action}), it would be
possible to describe a system of having the emergent conformal symmetry approximately, where the gravity scalar field
is dynamical and driven by the conformal anomaly, left hand side of Eq.(\ref{GS-eq}).}

Due to the presence of the cosmic scalar potential $V(\Phi)$, the equations (\ref{S-eq},\ref{G-eq}) do not allow the
exact flat spacetime solution even for a constant gravity scalar field in vacuum. However, in a non-cosmic scale, it is
found that the deviation from the flat spacetime could be negligible provided that the cosmic potential satisfies a
certain condition.

\section{The Cosmological Evolution}

We investigate the cosmological evolution, assuming a homogeneous gravity scalar field $\Phi(t)$ depending only on
time, and adopting the Robertson-Walker metric with a vanishing spatial curvature($k=0$),
\begin{eqnarray}
ds^{2}=dt^{2}-S^{2}(t)[dr^{2}+r^{2}d\Omega^{2}]. \label{metric}
\end{eqnarray}

Then, Eqs.(\ref{G-eq},\ref{GS-eq}) are reduced to a set of cosmological equations as shown below. Denoting the time
derivative as an over-dot,
\begin{eqnarray}
3\Phi^{2}H^{2} +6H\Phi\dot{\Phi} +3\dot{\Phi}^{2} = \rho_{m} + \rho_{\Lambda}(\Phi) , \label{CosG00-eq}
\end{eqnarray}
\begin{eqnarray}
2 a \Phi^{2} +H^{2}\Phi^{2} +4H\Phi\dot{\Phi} +2\Phi\ddot{\Phi} - \dot{\Phi}^{2} = -p_{m} + \rho_{\Lambda}(\Phi) ,
\label{CosG11-eq}
\end{eqnarray}
\begin{eqnarray}
\rho_{m} - 3p_{m} +4\rho_{vac} +4V(\Phi) = \Phi\frac{\partial V(\Phi)}{\partial \Phi} , \label{CosD-eq}
\end{eqnarray}
where the Hubble parameter $H \equiv \dot{S}/S$, the acceleration parameter $~a \equiv \ddot{S}/S$. With the matter
pressure $p_{m}$, the matter density $\rho_{m}$, $~T^{(m)0}_{0}=\rho_{m}+\rho_{vac}$,
$~T^{(m)1}_{1}=T^{(m)2}_{2}=T^{(m)3}_{3}=-p_{m}+\rho_{vac}$, where $\rho_{vac}$ is the vacuum energy which could be
generated by quantum effects in the matter Lagrangian. We can see that the sum of the vacuum energy $\rho_{vac}$ in the
matter Lagrangian and the cosmic potential $V(\Phi)$ plays the role of the total dark energy density,
$\rho_{\Lambda}(\Phi) \equiv \rho_{vac} + V(\Phi)$.

\section{The Cosmic Potential to Regulate the Dark Energy}

We try to find a cosmic potential in the conformally coupled gravity which can regulate the vacuum energy in the matter
Lagrangian. For a given vacuum energy $\rho_{vac}$ with the matter density $\rho_{m}$ and the matter pressure $p_{m}$
from the matter Lagrangian, a gravity scalar field $\Phi$ would satisfy Eq.(\ref{CosD-eq}). For a different vacuum
energy $\rho'_{vac}$, the equation (\ref{CosD-eq}) would be satisfied with a different gravity scalar field $\Phi'$
instead of $\Phi$. For these two different vacuum energies $\rho_{vac}$ and $\rho'_{vac}$, the difference in the total
dark energy density is found as
\begin{eqnarray}
\Delta\rho_{\Lambda} =\rho_{\Lambda}(\Phi') -\rho_{\Lambda}(\Phi) = V(\Phi')-V(\Phi)+\rho'_{vac}-\rho_{vac} =
\frac{1}{4} \left[ \Phi'\frac{\partial V(\Phi)}{\partial \Phi}\mid_{\Phi'} - \Phi\frac{\partial V(\Phi)}{\partial
\Phi}\mid_{\Phi} \right] . \label{Zero-eq}
\end{eqnarray}

Therefore, we can have the same total dark energy density $\rho_{\Lambda}$ in the conformally coupled gravity whatever
the vacuum energy $\rho_{vac}$ is in matter Lagrangian provided that the cosmic scalar potential satisfies the
equation, $\Phi\partial V(\Phi)/\partial\Phi = \alpha$ for a constant $\alpha$. Thus we choose the cosmic scalar
potential quite uniquely as;
\begin{eqnarray}
V(\Phi) = \alpha\ln\frac{\Phi}{\mu} , \quad \rho_{\Lambda}(\Phi) = \rho_{vac} + \alpha\ln\frac{\Phi}{\mu}
\label{Pot-eq} ,
\end{eqnarray}
where $\alpha$ is a dimensionful constant which would have some other physical origin. Both the dimensionful constant
$\mu$ and the vacuum energy $\rho_{vac}$ in matter Lagrangian hide into the total dark energy density
$\rho_{\Lambda}(\Phi)$ and do not appear as independent physical quantities at all.

For the cosmic potential Eq.(\ref{Pot-eq}), the total dark energy density $\rho_{\Lambda}$ satisfies the constraint
equation (\ref{CosD-eq}) with the matter density $\rho_{m}$ and the matter pressure $p_{m}$,
\begin{eqnarray}
\rho_{m} - 3p_{m} + 4\rho_{\Lambda}(\Phi) = \alpha . \label{CR-eq}
\end{eqnarray}
The left hand side of this equation is the trace of the total energy momentum tensor including all matter and total
dark energy of our universe. If we interpret Eq.(\ref{CR-eq}) as the trace anomaly cancellation equation which states
that the total trace anomaly of our universe vanishes, then the constant $-\alpha$ in Eq.(\ref{CR-eq}) would be the
conformal anomaly which is generated in matter Lagrangian by quantum effects. Thus, we may speculate that the cosmic
potential (\ref{Pot-eq}) is an effective potential to describe the conformal anomaly generated in matter Lagrangian by
quantum effects. It is found that the cosmic potential (\ref{Pot-eq}) is the linear form of coupling
$\phi\Theta^{\mu}_{\mu}$ between a scalar field $\phi$ and a conformal anomaly $\Theta^{\mu}_{\mu}$, which might have
an origin in quantum chromodynamics, considered in \cite{No-go,anomal} with $\Phi \equiv e^{-\phi}/\kappa_{0}$. Thus,
we can say $\alpha$ as a conformal anomaly parameter.

Taking a time derivative Eq.(\ref{CosG00-eq}), and using Eqs.(\ref{CosG11-eq},\ref{CosD-eq},\ref{Pot-eq}), we have the
matter energy conservation equation,
\begin{eqnarray}
\frac{d}{dt}\rho_{m} + 3H(\rho_{m}+p_{m}) = 0 , \label{matter-eq}
\end{eqnarray}
which is independent of the gravity scalar field $\Phi$.

\section{The Late Time Cosmology of Our Universe}

In the late time evolution of our universe dominated by cold matter with the matter pressure $p_{m}=0$, the matter
energy conservation equation (\ref{matter-eq}) tells $\rho_{m} = A/S^{3}(t)$ with a constant $A$ \cite{LCDM}. Taking a
time derivative of equation (\ref{CR-eq}) with the help of the explicit form Eq.(\ref{Pot-eq}), we find that
\begin{eqnarray}
\frac{d \Phi}{dt} = -\frac{\Phi}{4\alpha} \frac{d \rho_{m}}{dt} = \frac{3}{4\alpha}H\Phi \rho_{m} \label{late-Phi} .
\end{eqnarray}
Thus, we can find the solution of the gravity scalar $\Phi(t)$ and the corresponding gravitational constant $8\pi
G=1/\Phi^{2}(t)$ as a function of the scale factor $S(t)$,
\begin{eqnarray}
\Phi(S) = \Phi_{\infty}e^{-\frac{A}{4\alpha}\frac{1}{S^{3}(t)}}, \quad G(S) =
G_{\infty}e^{\frac{A}{2\alpha}\frac{1}{S^{3}(t)}} \label{late-G-sol} ,
\end{eqnarray}
where $\Phi_{\infty}$ and $G_{\infty}$ are the values of the gravity scalar field and the gravitational constant in the
infinite future, respectively.

The relative variations of the gravity scalar field $\Phi$ and the gravitational constant $G$ in the matter dominated
universe are determined by the matter density ratio $\rho_{m}/\alpha$ as
\begin{eqnarray}
\frac{\dot{\Phi}}{H\Phi} = \frac{3}{4}\frac{\rho_{m}}{\alpha} , \quad \frac{\dot{G}}{H G} =
-\frac{3}{2}\frac{\rho_{m}}{\alpha} \label{rates} .
\end{eqnarray}

However, $\Phi$ should be a constant in the radiation dominated early universe from Eq.(\ref{CR-eq}) observing
$p_{m,rad}=\rho_{m,rad}/3$ for relativistic matter. Therefore, this conformally coupled gravity model is reduced to
General Relativity giving the Friedmann equations with a constant gravitational constant $G_{rad}=G_{\infty}$ and a
constant dark energy density $\rho_{\Lambda_{rad}}=\alpha/4$ in the radiation dominated early universe.

Inserting Eq.(\ref{late-Phi}) into the cosmological equations (\ref{CosG00-eq},\ref{CosG11-eq}), we have a set of exact
cosmological evolution equations in the matter dominated late time universe as follows,
\begin{eqnarray}
3\Phi^{2}H^{2} = \alpha\frac{4(1+3x)}{~(4+3x)^{2}} ,\label{H2-eq}
\end{eqnarray}
\begin{eqnarray}
6\Phi^{2}a = \alpha \frac{3(1-x)(4+3x)^{2} - (1+3x)(16-48x+9x^{2})}{(4+3x)^{3}} , \label{a-eq}
\end{eqnarray}
where we have defined $x \equiv \rho_{m}/\alpha$, and used $\rho_{\Lambda}(\Phi)=(1-x)\alpha/4$ from Eq.(\ref{CR-eq})
with $p_{m}=0$. The cosmological equations (\ref{H2-eq},\ref{a-eq}) approximate to the Friedmann equations only if $x$
is much smaller than $1$, i.e. $x \ll 1$.

The jerk parameter which measures time variation of the cosmic acceleration parameter $a$ is obtained as
\begin{eqnarray}
j \equiv \frac{d^{3}S(t)}{S(t)dt} = \frac{\dot{a}}{H^{3}} +\frac{a}{H^{2}} = 1 -
\frac{9}{16}x\frac{64-32x+3x^{2}}{4+3x} + \frac{9}{8}x\frac{4+3x}{1+3x} + \frac{9}{4}\frac{a}{H^{2}}x\frac{8-3x}{4+3x}
\label{j-eq} ,
\end{eqnarray}
which depends only on $x$ because $a/H^{2}$ is a function of $x$ only from Eqs.(\ref{H2-eq},\ref{a-eq}).

\section{Implications on the Cosmology in the Present Universe}

Because the matter in our universe is believed to be cold with $p_{m}=0$ at present \cite{LCDM}, the matter density
$\rho_{m}$ and the current dark energy density $\rho_{\Lambda}$ would satisfy
\begin{eqnarray}
\rho_{m} +4\rho_{\Lambda} = \alpha \equiv 4\rho_{\Lambda_{\infty}} \label{CR}.
\end{eqnarray}
As our universe expands, the matter density will eventually die out to zero. Thus, the total dark energy density
$\rho_{\Lambda_{\infty}}=\alpha/4$ will prevail our universe in the infinite future. $\rho_{\Lambda_{\infty}}$ is the
very dark energy density $\rho_{\Lambda_{rad}}$ in the radiation dominated early universe with
$p_{m,rad}=\rho_{m,rad}/3$.

Recently, the ratio of the cosmic acceleration parameter $a$ to the Hubble parameter $H^{2}$ was determined through the
help of Type-I supernovae as standard candles \cite{ACC1,ACC2,LCDM},
\begin{eqnarray}
\left(\frac{a}{H^{2}}\right)_{exp} \simeq 0.55 \label{ratio-exp}.
\end{eqnarray}
From the ratio of Eq.(\ref{a-eq}) to Eq.(\ref{H2-eq}), $\frac{a}{H^{2}}$, which is a monotonically decreasing function
of $x$, we can find a single solution $x_{exp}$ which satisfies the experimental ratio $(a/H^{2})_{exp}$ numerically.
If we use the experimental value (\ref{ratio-exp}), then we find the corresponding solution $x_{exp}=0.476$. For this
$x_{exp}$, the cosmological equations (\ref{H2-eq},\ref{a-eq}) do not approximate to the Friedmann equations because
the higher $x$ terms are not negligible. This $x_{exp}=0.476$ determines the matter density $\rho_{m}$, the dark energy
density $\rho_{\Lambda}$, the jerk parameter $j$, the critical density $\rho_{c}$ at present, respectively, as
\begin{eqnarray}
\rho_{m}=0.476\alpha, \quad \rho_{\Lambda}=0.131\alpha, \quad j= 0.47, \quad \rho_{c} \equiv 3\Phi^{2}H^{2} =
\frac{3}{8\pi G}H^{2} =0.33\alpha  \label{fit-exp1}.
\end{eqnarray}
Thus, the ratio of the dark energy density to the matter density in the present universe is $\rho_{\Lambda}/\rho_{m}
\cong 1/4$, which is quite smaller than the value $\rho_{\Lambda}/\rho_{m} \cong 7/3$ determined by the Friedmann
equations based on the same data (\ref{ratio-exp}). The jerk parameter $j=0.47$ predicted by this model is quite
definite as General Relative predicts $j=1$.

Because only about $4\%$ of the critical density $\rho_{c} = (\frac{3}{8\pi G}H^{2})_{exp}$ is the ordinary baryonic
matter \cite{LCDM}, the baryonic matter density is estimated as $\rho_{baryon} \simeq 0.013 \alpha$. Thus, the dark
matter density is $\rho_{DM} \simeq 0.463\alpha$, which dominates all matter density $\rho_{m}$ as
$\rho_{DM}/\rho_{m}=0.97$ in this model.

From the difference between the infinite future and the current total dark energy densities, we can estimate the
increase of gravity scalar field $\Phi_{\infty}^{2} \simeq 1.27\Phi^{2}_{now}$ and the corresponding decrease of
gravitational constant $G_{N\infty} \simeq 0.79 G_{N,now}$ in the infinite future, restoring to the gravitational
constant in the radiation dominated early universe.

\section{Conclusions}

We have proposed the conformally coupled gravity with the uniquely determined cosmic potential, which describes a
system of having the emergent conformal symmetry. We have found that the dark energy density is regulated to the order
of the conformal anomaly parameter. The logarithmic cosmic potential might be an effective way to describe the
conformal anomaly which is generated by quantum effects in the matter Lagrangian. The conformal anomaly parameter
$\alpha$ introduced in the cosmic potential may provide a new fundamental length scale of gravity instead of the
Newton's constant as $\alpha \sim (10^{-3}eV)^{4} \sim (0.1mm)^{4}$, which is an order of the matter density in our
present universe. The smallness of the conformal parameter $\alpha$ would ensure that the local gravity in a solar
system scale is not affected by the presence of the cosmic potential.

In the matter dominated late time universe, we have a set of exact cosmological equations which deviate from the
Friedmann equations significantly. However, in the radiation dominated early universe, we regain the Friedmann
equations of General Relativity having a constant gravity scalar field. Thus, the conformally coupled gravity with the
logarithmic cosmic potential is a minimal extension of General Relativity to implement the dark energy regulation
mechanism, replacing the dimensionful Newton's constant with the gravity scalar field.

Based on the recent observational cosmic expansion data for $(a/H^{2})_{exp}$, we have found that the dark energy
density is about 1/4 of the matter density in the present universe, which is quite smaller than the value $7/3$
determined by the Friedmann equations. As a characteristic prediction of this model, the jerk parameter at present has
a quite definite value $j=0.47$ as General Relativity predicts $j=1$.

Since the regulation mechanism of the dark energy, through the emergent conformal symmetry in the conformally coupled
gravity with the unique cosmic potential, may resolve the puzzles for the dark energy, we hope that the jerk parameter
$j$ will be measured precisely to test this model in near future.




\begin{thebibliography}{99}


\bibitem{ACC1} A.G. Riess {\it et al.}, Astron. J. {\bf 116}, 1009 (1998); Astron. J. {\bf 118}, 2668
(1999); Astron. J. {\bf 607}, 665 (2004); S. Perlmutter {\it et al.}, Nature {\bf 398}, 51 (1998); Astron.
J. {\bf 517}, 565 (1999); R. Knop {\it et al.}, Astron. J. {\bf 598}, 102 (2003); B. Barris {\it et al.},
Astron. J. {\bf 602}, 571 (2004);

\bibitem{ACC2} S. Perlmutter {\it et al.}, Astrophys. J. {\bf 517}, 565 (1999).

\bibitem{I1} A. Silvestri and M. Trodden, Rept. Prog. Theor. {\bf 72}, 096901 (2009).

\bibitem{I2} M. Sami, arXiv:0904.3445, (2009).

\bibitem{Higgs} The CMS Collaboration, Phys. Lett. B {\bf 716}, 30 (2012).

\bibitem{BD1} C.H. Brans and R.H. Dicke, Phys. Rev. {\bf 124}, 925 (1961).

\bibitem{BD2} P.G. Bergmann, Int. J. Theor. Phys. {\bf 1}, 25 (1968).

\bibitem{IGO1} A. Zee, Phys. Rev. Lett. {\bf 42}, 417 (1979).

\bibitem{IGO2} L. Smolin, Nucl. Phys. {\bf B160}, 253 (1979).

\bibitem{IGO3} A. Adler, Rev. Mod. Phys. {\bf 54}, 729 (1982).

\bibitem{IGO4} A.D. Sakharov. Dokl. Akad. Nauk. SSSR 117, 70 (1967) [Sov. Phys. Dokl. {\bf 12}, 1040
(1967)].

\bibitem{IGO5} P. Batra, K. Hinterbichler, L. Hui, and D. Kabat, Phys. Rev. {\bf D78}, 043507 (2008).

\bibitem{IG0} T. Padmanabhan, {\it Quantum Conformal Fluctuations, Induced Gravity And Cosmology}, CERN
Preprint:CERN-TH-3706 (Sep. 1983).

\bibitem{Odint} I.L. Buchbinder and S.D. Odintsov, Int. J. Mod. Phys. A {\bf 3}, 1859 (1988); Yad.Fiz. {\bf 46}, 1233
(1987).

\bibitem{IG1} F.S. Accetta, D.J. Zoller, and M.S. Turner, Phys. Rev. {\bf D31}, 3046 (1985).

\bibitem{IG2} F.S. Accetta and J.J. Trester, Phys. Rev. {\bf D39}, 2854 (1989).

\bibitem{IG3} E. Carugno, S. Capozziello, and F. Occhionero, Phys. Rev. {\bf D47}, 4261 (1993).

\bibitem{IG4} D. La, Phys. Rev. {\bf D44}, 1680 (1991); {\it The "Scaled" Induced-Gravity Inflationary
Cosmology}, U.C. Berkely Preprint:CfPA-TH-90-015A (Jul. 1990).

\bibitem{IG5} J.L. Cervantes-Cota and H. Dehnen, Nucl. Phys. B {\bf 442}, 391 (1995).

\bibitem{IG6} W.F. Kao, Phys. Lett. A {\bf 147}, 165 (1990).

\bibitem{IG7} R. Fakir and  W.G. Unruh, Phys. Rev. D {\bf 41}, 1792 (1990).

\bibitem{IG8} D.I. Kaiser, Phys. Rev. {\bf D49}, 6347 (1994).

\bibitem{IG9} V. Faraoni, Class. Quant. Grav. {\bf 26}, 145014 (2009).

\bibitem{No-go} S. Weinberg, Rev. Mod. Phys. {\bf 61}, 1 (1989).

\bibitem{YP2} Y. Yoon, Phys. Rev. {\bf D59}, 127501 (1999); C.J. Park and Y. Yoon, Gen. Rel. Grav. {\bf 29}, 765
(1997).

\bibitem{anomal} R.D. Peccei, J. Sol$\grave{a}$, and C. Wetterich, Phys. Lett. B {\bf 195}, 183 (1987).

\bibitem{LCDM} S.M. Carroll, Liv. Rev. Rel. {\bf 4}, 1 (2001).

\end{thebibliography}
\end{document}